# Phase Modulators Based on High Mobility Ambipolar ReSe$_2$ Field-Effect Transistors

Nihar R. Pradhan[1,2], Carlos Garcia[2,3], Bridget Isenberg[2,4], Daniel Rhodes[2,3], Simin Feng[5], Shahriar Memaran[2,3], Yan Xin[2], Amber McCreary[6], Angela R. Hight Walker[6], Aldo Raeliarijaona[7], Humberto Terrones[7], Mauricio Terrones[5,8,9,10], Stephen McGill[2], & Luis Balicas[2]

**We fabricated ambipolar field-effect transistors (FETs) from multi-layered triclinic ReSe$_2$, mechanically exfoliated onto a SiO$_2$ layer grown on *p*-doped Si. In contrast to previous reports on thin layers (~2 to 3 layers), we extract field-effect carrier mobilities in excess of $10^2$ cm$^2$/Vs at room temperature in crystals with nearly ~10 atomic layers. These thicker FETs also show nearly zero threshold gate voltage for conduction and high ON to OFF current ratios when compared to the FETs built from thinner layers. We also demonstrate that it is possible to utilize this ambipolarity to fabricate logical elements or digital synthesizers. For instance, we demonstrate that one can produce simple, gate-voltage tunable phase modulators with the ability to shift the phase of the input signal by either 90° or nearly 180°. Given that it is possible to engineer these same elements with improved architectures, for example on *h*-BN in order to decrease the threshold gate voltage and increase the carrier mobilities, it is possible to improve their characteristics in order to engineer ultra-thin layered logic elements based on ReSe$_2$.**

Layered rhenium-based transition metal dichalcogenides (TMDCs), or ReX$_2$ where X = S, Se) are the subject of a renewed interest due to their unique anisotropic optoelectronic properties [1–6]. Due to a lattice distortion these materials crystallize in a distorted triclinic-1*T*'-phase instead of the more conventional trigonal prismatic, or 2*H*-phase, and the rhombohedral or 3*R*-phase. The crystal structure of ReX$_2$ is special due to their in-plane motif, i.e. four Re atoms are arranged in a diamond-like shape with these diamonds forming atomic chains along the *b*-direction[7].

[1]Department of Chemistry, Physics and Atmospheric Sciences, Jackson State University, Jackson, MS 39217, USA. [2]National High Magnetic Field Laboratory, Florida State University, Tallahassee, FL 32310, USA. [3]Department of Physics, Florida State University, Tallahassee, FL 32306, USA. [4]Lincoln High School, Tallahassee, FL 32311, USA. [5]Department of Physics and Center for 2-Dimensional and Layered Materials, The Pennsylvania State University, University Park, PA 16802, USA. [6]Engineering Physics Division, Physical Measurement Laboratory, NIST, Gaithersburg, Maryland 20899, USA. [7] Rensselaer Polytechnic Institute, Department of Physics, Applied Physics, and Astronomy, Troy, NY 12180 USA. [8]Department of Materials Science & Engineering, The Pennsylvania State University, University Park, PA 16802, USA. [9]Department of Chemistry, The Pennsylvania State University, University Park, PA 16802, USA. [10]Institute of Carbon Science and Technology, Faculty of Engineering, Shinshu University, Nagano 380-8553, Japan. Correspondence should be addressed either to N.R.P. (nihar.r.pradhan@jsums.edu) or to L. B. (balicas@magnet.fsu.edu)

In contrast to other more intensively studied layered dichalcogenides such as $Mo(S,Se)_2$ or $W(S,Se)_2$, which display a transition from an indirect to a direct band gap when exfoliated down to the monolayer limit[8], the Re-based TMDCs show nearly layer-independent ($ReS_2$)[9], or very weakly layer-dependent ($ReSe_2$), optical and vibrational properties[6]. This has been interpreted as evidence for an extremely weak inter-planar coupling although angle resolved photoemission spectroscopy observes an out-of-plane electronic dispersion, indicating that in fact the interlayer coupling in $ReSe_2$ is appreciable[10].

However, their triclinic symmetry makes both compounds optically biaxial, resulting in an anisotropic planar response with respect to the optical polarization[3, 11, 12]. The Raman-active modes of thin layers of both compounds have also been found to be anisotropic[13-19].

According to Density Functional Theory (DFT) calculations, bulk and monolayer $ReS_2$ have nearly identical band structures with direct bandgaps of 1.35 eV and 1.44 eV, respectively[2]. These values are relatively close to those extracted from high resolution electron energy loss spectroscopy, which finds direct band gaps of 1.42 eV and of 1.52 eV for bulk and monolayer $ReS_2$, respectively[17]. In contrast, the DFT calculations in Ref.[19] indicate that $ReS_2$ displays an indirect band gap that is very close in energy with respect to the direct one in the bulk and also in monolayers. Photoemission in Ref.[10] suggests that $ReSe_2$ would also possess an indirect band gap at all thicknesses which is consistent with Local Density Approximation calculations revealing gaps of 0.92 eV for the bulk and 1.22 eV for monolayer.[3] However, photoluminescence measurements at $T$=10 K coupled to GW and Bethe-Salpeter calculations find that the bandgap of $ReSe_2$ increases as the number of layers decreases, displaying values of 1.37 eV for the bulk and of 1.50 eV for the monolayer, while maintaining a direct band gap and independence of the number of layers[3]. In comparison with the other TMDCs crystallizing in 2$H$-phase, a direct band gap approaching ~ 1.5 eV would make these compounds particularly appealing for photo-sensing and photovoltaic applications (according to the Shockley-Queisser limit). However, as seen from the above results, the nature of the band gap in ReSe$_2$, i.e. direct or indirect, remains unclear. To date, there are no reports on room temperature PL from ReSe$_2$. Despite our multiple attempts, we were also unable to collect room temperature PL data from this compound, although we can easily extract a PL signal from those compounds crystallizing in the 2$H$ structure, which are known to display an indirect band bap in the bulk[8]. This observation is particularly difficult to reconcile with a direct band gap.

Nevertheless, photodetectors[20] based on high-quality chemical vapor deposition grown ReS$_2$ were reported to yield photoresponsivities as high as $R \cong 604$ A/W corresponding to an enormous external quantum efficiency EQE = 1.50 x 10$^5$ % and a specific detectivity[21] $D^* \cong$ 4.44 x 10$^8$ m (Hz)$^{1/2}$/W. For ReS$_2$ stacked onto $h$-BN, values as high as $R$ = 88 600 A/W, EQE = 2 x 10$^7$ %, and $D^*$ = 1.182 × 10$^{10}$ m (Hz)$^{1/2}$/W were reported[22], indicating that the substrates play a significant role on the performance of these compounds. As for ReSe$_2$, values as high as $R$ = 3.68 × 10$^4$ A/W were reported after improving the resistance of the source contact via a triphenyl phosphine-based $n$-doping technique[23]. These pronounced photoresponsivities would suggest that the band gap of these materials might indeed be direct.

Here, we show that FETs based on multilayered triclinic ReSe$_2$ mechanically exfoliated onto SiO$_2$, display ambipolar behavior with very small threshold back-gate voltages, current ON to OFF ratios exceeding 10$^6$ and electron field-effect mobilities approaching 380 cm$^2$/Vs at room temperature. We find that Density Functional Theory calculations can replicate the observed Raman spectra for the bulk and for the monolayers concluding that all active Raman modes belong to the $A_g$ irreducible representation given that inversion is the only symmetry operation compatible with its structure. We also show that the ambipolarity of ReSe$_2$ opens up interesting opportunities for complementary logic electronics: for instance, we demonstrate that it is easy to produce very simple, gate-voltage controlled AC-voltage phase modulators as previously reported for ReS$_2$[24].

**Results and Discussion**

Figure 1a displays a scanning transmission electron microscopy (STEM) image of one of our exfoliated ReSe$_2$ single-crystals. The high crystallinity of these crystals is confirmed by the electron diffraction pattern collected along a direction perpendicular to the planes or along the [001] direction (see Figure 1b). Given its structural symmetry and similarity to ReS$_2$, ReSe$_2$ also tends to exfoliate in the form of nearly rectangular flakes[18,19], as seen in Figure 1c, which shows a micro-image of the ReSe$_2$ flake exfoliated onto a 285 nm thick SiO$_2$ layer grown on *p*-doped Si. As shown in the inset, according to atomic force microscopy (AFM) the thickness of the exfoliated crystal shown in 1c is approximately four atomic layers for an inter-planar lattice separation $c = 0.6702$ nm[5]. Figure 1c shows a micro-image of the ReSe$_2$ flake exfoliated onto a 285 nm thick SiO$_2$ layer grown on *p*-doped Si. Figure 1d shows the same crystal after the deposition of the electrical contacts, i.e. 50 nm of Au on 5 nm of Cr. The electrical contacts were deposited using standard e-beam lithography and e-beam evaporation techniques. This configuration of six contacts allows one to measure the Hall-effect to extract the Hall-mobilities which will be reported elsewhere. Figures 1e and 1f display the experimental Raman scattering spectra for a monolayer and a five layer crystal, respectively. Their expected theoretical spectra, from which we index the peaks in Figure 1e, are shown in Figure 1g. In order to compare with the experimental Raman results, ab-initio density functional theory (DFT) and density functional perturbation theory (DFPT) calculations were performed for monolayer and bulk ReSe$_2$ as implemented in the plane wave code CASTEP[25, 26]. The starting structure for the bulk crystal was obtained from Lamfers *et al.*[27]. Monolayer, few-layer and the bulk crystal exhibit just inversion symmetry and belong to the $P\bar{1}$ space group. Local density approximation (LDA) using the Ceperly-Alder-Perdew and Zunger (CA-PZ) functional[28, 29] with $6 \times 6 \times 1$ Monkhorst-Pack *K*-points and a plane waves cut-off of 440 eV with a norm-conserving pseudopotential was implemented in the calculations. The structures were relaxed, including the unit cells, until the forces became smaller than 0.01 eV/Å and with self-consistent energy tolerances inferior to $5\times10^{-7}$ eV/atom. For the monolayer case a vacuum of 21 Å between the layers was considered. Due to fact that the only symmetry operation present in the monolayer, few-layers, and bulk, is inversion symmetry, there is just one Raman active irreducible representation, i.e. $A_g$. Thus, in Figure 1e we have labeled the Raman peaks in a sequence from low to high frequencies as $A_g$ with increasing exponent number. In the Supplementary Information, we included a table, i.e. Table S1, which provides the calculated phonon frequencies for all the bulk and monolayer Raman modes.

Figure 2 presents the overall electrical response of two field-effect transistors (FETs) built from exfoliated flakes composed of 4 and 10 layers, respectively. We observed a large variability in the response of the characterized FETs, with thicker crystals displaying considerably larger field-effect mobilities and smaller hysteresis and threshold gate-voltages relative to thinner ones as illustrated by both examples in Figure 2 (see also Figures S1 and S2 in Supplementary Information file for electrical data from other samples). The lower mobilities and poorer overall electrical performance of FETs built from thinner $ReSe_2$ crystals has been widely observed in transition metal dichalcogenides [30-32] and attributed to a more pronounced role for Coulomb scattering from the impurities at the interface with the $SiO_2$ layer and particularly from the adsorbates accumulated at the top layer of the semiconducting channel[30-32]. In thicker crystals/flakes the top layers play the role of capping layers, protecting the middle layers that carry most of the electrical current, from oxidation and adsorbates, while the bottom layers can partially screen the spurious charges at the interface. Simulations have also shown that carrier mobilities peak for samples having approximately 10 layers[30]. Figure 2a displays the drain to source current $I_{ds}$, extracted under a bias voltage $V_{ds}$ = 50 mV, as a function of gate voltage $V_{bg}$ for a FET based on an $n$ = 10 layers crystal. One observes i) the near absence of hysteresis and ii) a threshold gate-voltage $V_{th} \cong \pm 40$ V beyond which conduction occurs at room temperature. Notice also the ambipolar behavior, or electron and hole-conduction, with ON to OFF current ratios of nearly ~$10^7$ for electrons and of ~ $10^6$ for holes, albeit with poor subthreshold swings of typically ~3.5 V per decade. Previously ambipolarity was reported only for black phosphorus[24] and for MoSe$_2$[49]. Hence, ReSe$_2$ becomes the third 2D material to display ambipolarity in absence of ionic liquid gating or dielectric, heterostructure or contact engineering, thus displaying a potential for applications in in complementary logic electronics.

Figure 2b shows $I_{ds}$ as a function of $V_{bg}$ for the same sample but for several temperatures $T$. As $T$ is lowered, one needs to reach progressively higher threshold gate voltages $V^t_{bg}$ to observe carrier conduction. We have observed this increase in $V^t_{bg}$ in most of the TMDs we have measured, ascribing it to a combination of factors, such as disorder-induced localization at the interface with the SiO$_2$ layer, and the role of the Schottky barriers at the electrical contacts[33]. Figures 2c and 2d illustrate a comparison between 2- and 4-terminal measurements performed in a FET based on a ReSe$_2$ crystal with $n$ = 4 layers. Here, 2-terminal measurements indicate that current flows through the source and drain contacts which are also used for sensing the voltage. The overall response of this FET is noticeably inferior with respect to the $n$ = 10 one: under the same bias voltage $V_{ds}$ = 50 mV one extracts nearly 100 times less current leading to ON to OFF ratios reaching only $10^4$ and $10^3$ for holes and electrons, respectively. The rather large $V^t_{bg}$ of ~ +40 and ~ - 30 V at $T$ = 275 K for the $n \cong 4$ sample suggest that this sample is considerably more disordered than the $n \cong 10$ layers one: a large fraction of the initially accumulated carriers become trapped by defects and spurious charges in the material and at the interface. Figures 2e and 2f display $I_{ds}$ as a function of $V_{bg}$ for measurements based on 2- and 4-terminal configurations respectively, at several temperatures. Notice how the $V^t_{bg}$ are nearly $T$-independent, supporting the notion that they are associated with a constant number of defects in the material and/or with spurious charges at the interface.

Figure 3 displays the field-effect mobilities extracted from both samples ($n = 10$ and $n = 4$ layers) as a function of the temperature $T$, when using the conventional MOSFET transconductance formula, i.e. $\mu_{FE} = c_g^{-1}\, d\sigma/dV_{bg}$, where $\sigma = j_{ds}/E_{ds}$ is the conductivity and $c_g = e_r e_0/d = 12.116 \times 10^{-9}$ F/cm$^2$ is the gate capacitance (where $d = 285$ nm is the thickness of the SiO$_2$ layer). For the thicker $n = 10$ sample, we measured the FET response only via the 2-terminal configuration. For the thinner $n = 4$ sample, we measured the mobility using the 2- as well as the 4-terminal method, since the first one yielded unusually small mobilites when compared to those of the $n \cong 10$ samples. Our intention was to verify if this difference would be attributable to worse electrical contacts. Remarkably, for the $n = 10$ sample, whose data is shown in Fig. 3a, we observed nearly temperature independent electron mobilities with values around $\sim 400$ cm$^2$/Vs. Meanwhile the hole mobilities varied significantly, increasing by more than one order of magnitude upon cooling, that is from $\sim 20$ cm$^2$/Vs at room temperature to $\sim 400$ cm$^2$/Vs at 75 K. Figure 3b displays the 2- as well as the 4-terminal electron and hole mobilities for $n = 4$ sample as a function of the temperature. Solid green and solid maroon dots depict hole and electron mobilities measured through a 2-terminal configuration. Open green and maroon dots depict hole and electron mobilities measured in a 4-terminal configuration, respectively. In the whole range of temperatures the mobilities of the $n = 4$ sample were considerably smaller than those of the $n = 10$ one, which displayed two-terminal hole mobilities on the order of just 1 cm$^2$/Vs and electron mobilities one order of magnitude smaller. These values for the 2-terminal mobility of the thinner sample were very similar to those reported by Zhang et al.[34]

The mobility values for the $n = 10$ sample were considerably higher than those previously reported for multilayered samples, which have been found to display electron-doped-like responses with two-terminal field-effect mobilities ranging from only ~1 to 10 cm$^2$/Vs[1,12,34-36]. In Figures S3, S4, and S5 (see Supplementary Information) we included data from a second multi-layered sample, i.e. $n = 8 - 9$ layers, which also displayed field-effect electron mobilities in excess of $10^2$ cm$^2$/Vs, with considerably smaller threshold gate-voltages relative to the thinner crystals. This indicates that thicker crystals indeed display higher mobilities and that this behavior is not confined to the sample shown here. In thinner samples, charge conduction tends to be dominated by higher contact resistances when measured in a 2-terminal configuration. Therefore, in order to extract the nearly intrinsic mobility of the $n = 4$ sample we re-measured it through a 4-terminal configuration. From these measurements, we obtained hole-mobilities of ~10 cm$^2$/Vs and an order of magnitude smaller electron mobilities at room temperature. This value for the 4-terminal hole-mobility was similar to the one reported by Zhang *et al.*[1] on the same material after transferring onto h-BN substrates. In contrast, our electron mobilities were similar to their values extracted from ReSe$_2$ FETs on SiO$_2$. Unsurprisingly, this indicates that the substrates, i.e. their roughness, presence of dangling bonds and of trapped charges affect the mobilities of thin ReSe$_2$ samples. Impurities should play a more predominant role in thinner crystals. The impurities to which we refer are those located at the interface with the SiO$_2$ layer as well as adsorbates on top of the channel resulting from air exposure during the fabrication process. Our observations would imply that previous reports underestimated the intrinsic performance of this compound. We discarded degradation under ambient conditions after evaluating the time dependence of the Raman signal, i.e. amplitude and width at half maximum of several of the observed Raman peaks as a function of time. We did not detect any deterioration in a time scale of a few days indicating that the previously discussed scattering mechanisms as well as the Schottky barriers at the electrical contacts are likely to be the main factors limiting the performance of ReSe$_2$ field-effect transistors. Notice that the mobility $\mu^h_{FE}$ of the holes in Figure 3a increased considerably as $T$ was lowered suggesting that it was phonon limited, or that phonons also played quite a relevant role at room temperature. In contrast, the electron and hole mobilities for the thin $n = 4$ layers sample showed quite different trends as a function of the temperature. The 2-terminal electron ($\mu^e_{2T}$) and hole ($\mu^h_{2T}$) mobilites decreased as a function of the temperature, indicating that the transport of charges was dominated by the Schottky barriers, or that the thermionic emission processes accross the contacts were suppressed at lower temperatures. This contrasting behavior also implied that the mobilities were sample dependent due in part to fluctuations in the quality of the contacts. The 4-terminal mobilities measured on the same device yielded a hole mobility ($\mu^h_{4T}$) that remained nearly constant at a value of ~20 cm$^2$/Vs as a function of the temperature. In contrast, the electron mobility ($\mu^e_{4T}$) increased from 0.3 cm$^2$/Vs at 300 K to 3 cm$^2$/Vs at 75 K. The previous report by Zhang *et al.*[1] also found nearly constant mobilities as a function of the temperature when h-BN was used as substrate.

To evaluate the quality of the electrical contacts we performed two-terminal measurements in the $n = 10$ sample to evaluate $I_{ds}$ as function of $V_{bg}$ under a fixed $V_{ds}$=50 mV at several temperatures, see Fig. 4. This evaluation was important since, as illustrated by Figure S1 (see Supplementary Information), the $I_{ds}$-$V_{ds}$ characteristics were non-linear confirming a prominent role for the Schottky barriers at the level of the electrical contacts with a concomitant performance loss in electrical transport in ReSe$_2$-based FETs. The transport of electrical charges across a Schottky barrier, resulting from the mismatch between the band structure of the metal and that of the two-dimensional material, is usually described in terms of the two dimensional thermionic emission equation[37-40]:

$$I_{ds} = AA^*T^n \left[-\frac{q\phi_{SB}}{k_B T}\right] \quad (1)$$

where $A$ is contact area of junction, $A^*$ is the two-dimensional equivalent Richardson constant, $n$ is an exponent acquiring a value of either 2 for a three dimensional semiconductor or of 3/2 for a two-dimensional one[40], $q = e$ is the electron charge, $\phi_{SB}$ is the Schottky barrier height, and $k_B$ is the Boltzmann constant. In order to evaluate $\phi_{SB}$ or the effective Schottky barrier at the contacts, in the top panel of Figures 4a and 4b we plotted the drain-source current $I_{ds}$ normalized by the power of the temperature $T^{3/2}$ as a function of $(q = e)/k_BT$ as obtained under several values of $V_{bg}$. Figure 4a corresponds to curves collected under $V_{bg} > 0$, while Figure 4b corresponds to curves measured under $V_{bg} < 0$. Red lines in both panels are linear fits from which we extracted the value of $\phi_{SB}$ ($V_{bg}$). Figure 4c displays the extracted values of $\phi_{SB}$ as a function of $V_{bg}$. The Schottky barrier height $\Phi_B$ for electrons and holes were extracted from $\phi_{SB}$ at large absolute values of the gate voltage (flat band condition indicated here by deviations to linear fits) yielding values of ~ 0.016 and 0.2 eV, respectively. These values must be contrasted with the work function W = 5.6 eV and the band gap of $\Delta$ = 1.19 eV reported for ReSe$_2$[12,41]. A Schottky barrier should be expected as the difference in energy between the work function of the deposited metal contacts Cr, or 4.5 eV, and the electron affinity $E_{EA} \cong (W - \Delta/2) \cong 5.005$ eV of ReSe$_2$, or $\Phi_B \cong +0.505$ eV. This value implies that Cr should pin the Fermi level within the conduction band of ReSe$_2$ thus explaining the rather small $\Phi_B$ ~ 0.015 eV extracted under positive gate voltages. The existence of a very small Schottky barrier could result from extrinsic factors like polymer residues resulting from the fabrication process. Remarkably, one also obtains a rather small Schottky barrier for holes of just $\Phi_B \cong 0.2$ eV, which is an unexpected result. Notice that a similar discrepancy was already observed by us for $\alpha$-MoTe$_2$[42]. Schottky barriers are likely the main factor limiting the hole-conduction in our ReSe$_2$ FETs while their asymmetry would explain the larger electron mobilities.

Transistors displaying ambipolar behavior could be useful for applications in telecommunications since they could simplify circuit design or improve the performance of signal processing. For instance, we demonstrate that the ambipolarity of ReSe$_2$ can be useful for the development of a phase shift modulator. For instance, Figure 5 displays the response of a $n = 4$ ReSe$_2$ based field-effect transistor, connected in series to a load resistor, upon the introduction of a sinusoidal modulation superimposed on its back-gate voltage, which we rename as the input-voltage, or $V_{in} = V_{bg} + V_{ac}$ ($\cong 1.5$ V). The readout oscillatory voltage $V_{out}$ is collected at a point located between the load-resistor, in this case $R_{load} = 100$ k$\Omega$, to which we apply a load voltage $V_{dd} = 50$ mV with respect to the ground, and the FET (see schematic of the circuit in Figure 5a). Figure 5a also displays $I_{ds}$ as a function of $V_{bg}$ where we placed three magenta squares indicating the constant values of $V_{bg}$ upon which oscillatory $V_{ac}$ signals were superimposed while the corresponding $V_{out}$ were collected. Figure 5b displays the phase shift of the $V_{out}$ signal relative to $V_{in}$ collected with a Lock-In amplifier as the gate voltage was swept from negative to positive values. Similarly to ambipolar $\alpha$-MoTe$_2$[42], the relative phase between both signals was observed to shift from ~0° for $V_{bg} < 0$ V, to ~90° for $0 \leq V_{bg} \leq 20$ V, and finally nearly inverted to ~170° for $V_{bg} > 40$ V. These phase shifts are better illustrated by the raw oscillatory signals observed and collected with an oscilloscope as shown in Figures 5c through 5e. When a negative $V_{bg}$ was applied to the back gate, $I_{ds}$ increased or decreased asynchronously with $V_{in}$, and consequently the corresponding $V_{out}$ also oscillated but in this case synchronously with $V_{in}$. This configuration corresponds to the so-called common drain mode and is illustrated by Figure 5c. In contrast, when a positive $V_{bg}$ was applied, the corresponding $V_{out}$ also oscillated although asynchronously, that is with a phase difference of nearly 180° with respect to $V_{in}$, as shown in Figure 5e (i.e. common-source mode). Remarkably, we observed a phase shift of ~ $\pi/2$ for $V_{bg} = 0$ V which we attribute to the very high impedance of the FET for gate voltages inferior to the respective threshold gate voltages for conduction. The lack of a sizeable conductivity, or of a real component in the FET impedance, implies that its impedance is dominated by an imaginary component associated with, for example, the gate capacitance or capacitive and/or inductive couplings at the level of the contacts. Since the frequency is the rate of change of the phase, phase modulators can be used for frequency modulation (FM), and in fact they are employed in commercial FM transmitters. In addition to a phase shift modulator, the ambipolarity of ReSe$_2$ can also be useful for the development of static voltage inverters, for example, by combining a ReSe$_2$-based FET gated to display *p*-type behavior with another one gated to behave as *n*-type[2, 44 – 46]. In supplementary Figure S6 we included the phase-shift as a function of the gate voltage for a second sample having approximately 10 layers. Hence, this behavior is reproducible among samples having a different number of layers.

**Conclusions**

In conclusion, given that the only symmetry operation present in the monolayer, few-layer and bulk ReSe$_2$ was inversion symmetry, our Raman study coupled to density functional theory calculations indicated that their Raman spectra contained only modes belonging to the $A_g$ irreducible representation. In addition and also in contrast to our previous studies on the isostructural ReS$_2$ compound[18], which was found to behave as an electron doped material, ReSe$_2$ displayed ambipolar behavior when contacted with Cr:Au electrodes. Relative to ReS$_2$, we observed a considerably larger variability in the response of field-effect transistors fabricated from few layers of ReSe$_2$ mechanically exfoliated onto SiO$_2$. FETs based on ~ 10 layers of ReSe$_2$ were observed to display up to one order of magnitude larger room temperature electron mobilities relative to FETs based on thinner flakes or on ReS$_2$, with, remarkably, negligible threshold gate voltage for carrier conduction. This suggests that the material was intrinsically of high quality, or prone to a relative low density of defects. Given that Raman scattering as a function of time indicated that ReSe$_2$ was rather stable under ambient conditions, the relatively poor performance observed in FETs fabricated from samples composed of just 3 to 4 layers was attributable to a poorer quality of the electrical contacts and to a more prominent role for impurity scattering from interfacial charges and adsorbates on the top layer. For instance, the exposure of the contact area to electron irradiation during the fabrication process is known to locally damage the material, for example, by inducing Se vacancies on the surface of the material[18,47]. But as discussed in Ref.[48] Se vacancies can induce a large amount of interfacial states within the band gap leading, according to the DFT calculations, to nearly complete Fermi level pinning and possibly to larger Schottky barriers. Radiation induced defects should be particularly detrimental to monolayers, with their role weakening as the surface to volume ratio decreases or as the sample thickness increases. This would contribute to explain the superior performance observed by us on FETs based on 8-10 layers when compared to those composed of 3-4 layers. It would also contribute to explain the relatively low mobilities previously reported by other groups for this compound[1,12].

Although our results point to considerably higher mobilities for the samples composed of $n = 10$ layers, one should take this observation with a grain of salt. For example, through a combination of measurements and simulations Das and Appenzeller concluded that four-terminal measurements would not be able to extract the intrinsic mobility of layered transition metal dichalcogenides given that both their carrier concentration and mobility become spatially dependent[49]. In addition one could also argue that we have not etched our crystals in a Hall bar geometry thus the metallic contacts deposited on the channel could affect its properties yielding incorrect values for its intrinsic mobility. However, we obtain comparable values for the 2- and the 4-terminal mobilities extracted for the $n = 4$ samples as well as similar values for 2- and 4- terminal mobilities for the $n = 10$ samples below $T \sim 100$ K. These observations strongly suggest that the higher mobilities for the $n = 10$ samples are intrinsic and do not a result from an artifact associated with the geometry or position of the contacts.

The ambipolarity of ReSe$_2$, when contrasted to the electron-doped behavior of ReS$_2$, bears resemblance with the $2H$-phase compounds MoSe$_2$ and MoS$_2$, where the former was reported by us as being ambipolar[50] while the second is well-known for behaving as electron doped. In TMDs, the nature of the carrier conduction, i.e. electron- or hole-like, is usually attributed to Fermi level pinning associated with the Schottky barriers around the metallic contacts[51]. However, it seems difficult to reconcile this scenario with the differences in crystallographic and electronic structures between all of these compounds. Instead, it suggests that the electron character of MoS$_2$ and ReS$_2$ is intrinsically associated with the density of sulphur vacancies[52]. In any case, as we showed here, the ambipolarity of TMDs like ReSe$_2$, allows one to produce quite simple logic elements having, for example, the ability to tune the phase of an incoming oscillatory signal towards 90º or 180º with the application of a single input voltage. It is therefore clear that these compounds have a remarkable potential for flexible logic applications. The current challenge is to understand and control the parameters limiting their performance, such as material quality, passivation, and Schottky barriers, in order to engineer commercial applications based on transition metal dichalcogenides.

**Materials and Methods**

**Crystal Synthesis.** ReSe$_2$ single crystals were synthesized through a chemical vapor transport (CVT) technique using either iodine or excess Se as the transport agent. Multi-layered flakes of ReSe$_2$ were exfoliated from these single crystals using the micromechanical cleavage technique and transferred onto *p*-doped Si wafers covered with a 285 nm thick layer of SiO$_2$.

**Characterization.** Atomic force microscopy (AFM) imaging was performed using the Asylum Research MFP-3D* AFM[†]. Raman spectra were acquired under ambient conditions using a micro-Raman spectrometer (Renishaw inVia micro-Raman[†]). A grating of 1800 lines/mm was used in the backscattering geometry, and a 100× objective lens was used to focus a laser spot size of ~1 μm onto the sample. The laser wavelength used to excite the samples was 514.5 nm (2.41 eV) from an Ar-Kr laser with a power around 0.1 mW to avoid any possible damage to the sample. Each Raman spectrum was measured with a 10 second accumulation time. Energy dispersive spectroscopy, to verify the stoichiometry, was performed through field-emission scanning electron microscopy (Zeiss 1540 XB).

**Trasmission Electron Microscopy.** Sub-Angstrom aberration corrected transmission electron microscopy was performed with a JEM-ARM200cF microscope.

**Device fabrication.** ReSe$_2$ crystals were mechanically exfoliated and then transferred onto a clean 285 nm thick SiO$_2$ layer. For making the electrical contacts 50 nm of Au was deposited onto a 5 nm layer of Cr via e-beam evaporation. Contacts were patterned using standard e-beam lithography techniques. After gold deposition, we proceeded with PMMA lift off in acetone. The devices were annealed at 300 °C for ~ 3 h in forming gas, followed by high vacuum annealing for 24 hours at 130 °C. Immediately after vacuum annealing, the devices were coated with a ~20 nm thick Cytop$^{TM}$ (amorphous fluoropolymer)$^†$ layer to prevent air exposure. Electrical characterization was performed by using a combination of a dual channel sourcemeters, Keithley 2400, 2612A and 2635 coupled to a Quantum Design Physical Property Measurement System.

**Data availability**. The datasets generated and analyzed during the current study are available from the corresponding author on reasonable request. A part of these data are included in this published article as Supplementary Information file.

$^†$Certain commercial equipment, instruments, or materials are identified in this manuscript in order to specify the experimental procedure adequately. Such identification is not intended to imply recommendation or endorsement by the National Institute of Standards and Technology, nor is it intended to imply that the materials or equipment are necessarily the best available for the purpose.

**Acknowledgements**

This work was supported by the Army Research Office through the MURI grant W911NF-11-1-0362. N.R.P. acknowledges NSF-HBCU-UP HRD-1332444 Institutional change through faculty advancement in instruction and mentoring-ICFAIM. L. B. also acknowledges support from the Office Naval Research DURIP Grant# 11997003. H.T. and A.R. are grateful to the National Science Foundation (EFRI-1433311), the Center for Computational Innovations (CCI) at Rensselaer Polytechnic Institute and the Extreme Science and Engineering Discovery Environment (XSEDE, project TG-DMR17008), which is supported by National Science Foundation grant number ACI-1053575. S.M. and C.G. acknowledge support from the NSF through DMR-1229217. The NHMFL is supported by NSF through NSF-DMR-1157490 and the State of Florida. A.M. and A.R.H.W. would like to acknowledge the NIST/National Research Council Postdoctoral Research Associateship Program and NIST-STRS for funding.

**Authors' Contributions**

NRP and LB conceived the project. DR synthesized the $ReSe_2$ single crystals. NRP and BI fabricated the field-effect transistor devices. NRP and BI measured temperature dependent transport properties and NRP analyzed the electrical transport data, CG, NRP, SM measured electrical properties for amplifier and inverter, SF, AM, AHW, MT performed Raman measurement. YX measured TEM and AR, HT contributed theoretical Raman calculation. NRP, SM and LB performed the electrical transport characterization. LB and NRP wrote the manuscript with the input of all co-authors.

**Additional Information**

**Supplementary information** accompanies this paper at https://doi.org/

**Competing Interests:** The authors declare no competing interests.

**Publisher's note:** Springer Nature remains neutral with regard to jurisdictional claims in published maps and institutional affiliations.

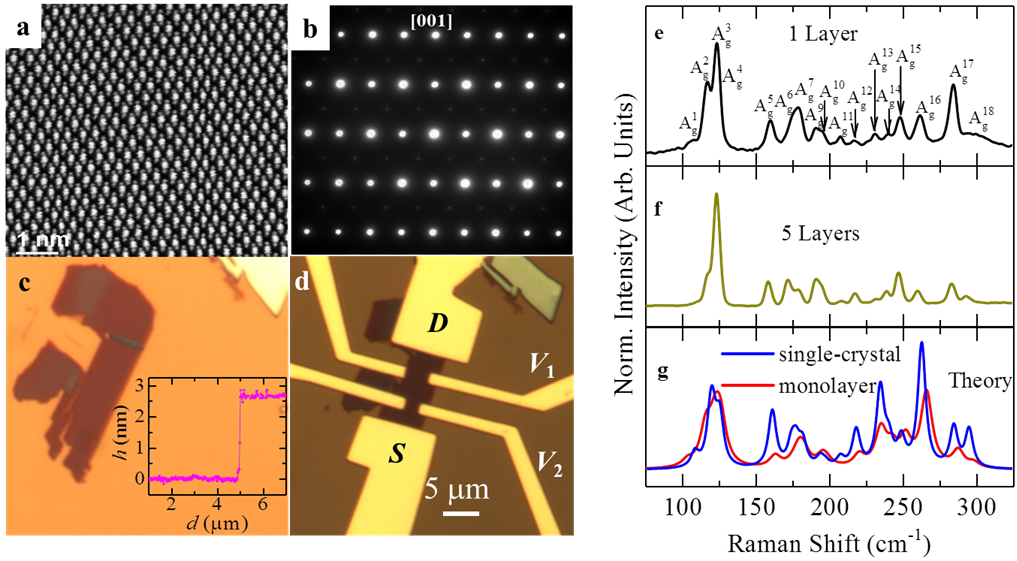

**Figure 1**. (**a**) Scanning transmission electron microscopy (STEM) image of an exfoliated $ReSe_2$ single-crystal displaying a chain-like atomic arrangement. (**b**) Electron diffraction pattern for a $ReSe_2$ single-crystal when the incident electron-beam is perpendicular to the planar atomic arrangement showing a rectangular planar Brillouin zone. (**c**) Micrograph of a typical few layered $ReSe_2$ single-crystal exfoliated onto $SiO_2$ which from AFM (inset) had a step height of 2.7 nm, or was four layers thick (for an inter-planar lattice separation $c = 0.6702$ nm [5]). (**d**) Micrograph of the same crystal after deposition of the electrical contacts which consisted of 50 nm of Au on a 5 nm layer of Cr. The larger electrical contacts were used to source (S) and drain (D) the current $I_{ds}$ when performing two-terminal measurements. The smaller contacts as $V_1$ and $V_2$ were used for voltage sensing in four-terminal measurements. For this sample, the separation between the current leads was $L \cong 10.5$ μm, the width of the channel was $w \cong 3.6$ μm and the separation between voltage leads was $l \cong 4.5$ μm. (**e**) Raman spectra of a $ReSe_2$ monolayer. Given that inversion symmetry was the only symmetry operation present in the monolayer, in few-layers and in the bulk, there was just one Raman active irreducible representation, i.e. $A_g$. Therefore, all peaks were associated with Raman $A_g$ modes. (**f**) Raman spectra for a crystal composed of five atomic layers. (**g**) Theoretical Raman spectra for monolayer (red) and bulk (blue) $ReSe_2$.

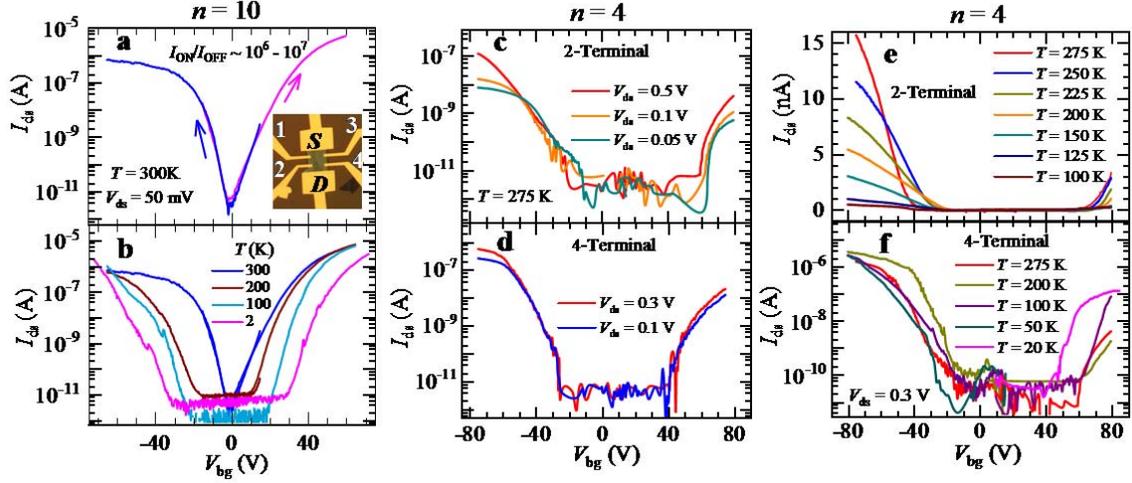

**Figure 2.** (**a**) Drain to source current $I_{ds}$ as a function of the gate-voltage $V_{bg}$ for a ten layer thick ReSe$_2$ crystal. Blue (magenta) markers depict decreasing (increasing) gate-voltage sweeps. Notice the near absence of hysteresis. Both traces were acquired at room temperature under a bias voltage of $V_{ds}$ = 50 mV. Inset: picture of the FET indicating the configuration of contacts. Source (S) and drain (D) contacts were used for two terminal measurements. The channel length and width of the device was 11.6 μm and 10.9 μm respectively. (**b**) $I_{ds}$ as a function of $V_{bg}$ for the same sample and for several temperatures ranging from $T$ = 300 K to 2 K. Notice the progressive emergence of a threshold gate-voltage which increased upon decreasing $T$. (**c**) Drain to source current $I_{ds}$ for a $n$ = 4 sample as a function of the back-gate voltage $V_{bg}$ in semi-logarithmic scale for several values of the bias voltage $V_{ds}$, measured at $T$ = 275 K through a two-terminal configuration. (**d**) Same as in (**c**) but measured *via* a four-terminal configuration. (**e**) $I_{ds}$ as a function of $V_{bg}$ for several temperatures measured via a two-terminal configuration in a linear scale. (**f**) Same as in (**e**) but in a logarithmic scale and measured through a four-terminal configuration. For both panels (e) and (f) a bias voltage $V_{ds}$ = 0.3 V was used.

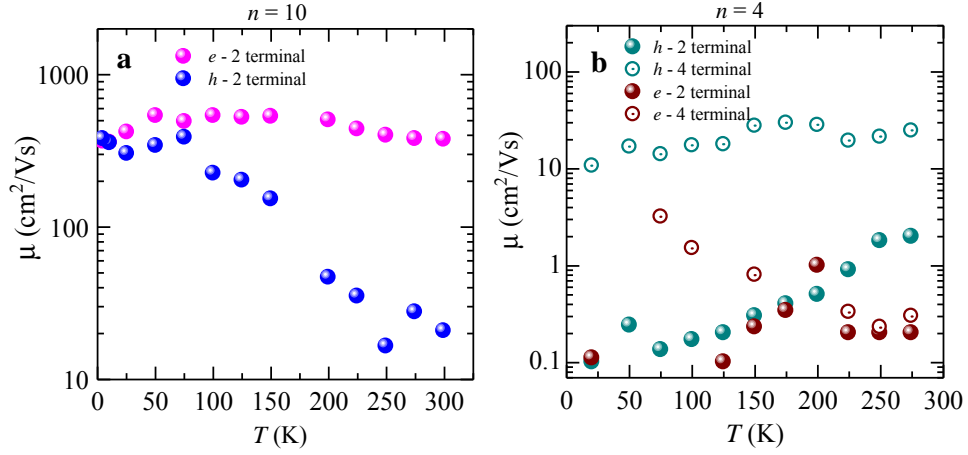

**Figure 3.** Electron- and hole- mobilities as a function of the temperature as extracted from the MOSFET transconductance formula for a (**a**) $n = 10$ and a (**b**) 4-layers thick sample. In (**a**) magenta and blue markers depict electron- and hole-mobilities respectively, as extracted from a two-terminal configuration under $V_{ds}= 50$ mV. In (**b**) dark cyan and brown markers depict hole- and electron mobilities, respectively. Solid and open circles indicate mobilities extracted from two- and four-terminal configurations, respectively. The drain to source voltage applied to the 4 layer sample was $V_{ds}= 0.3$ V.

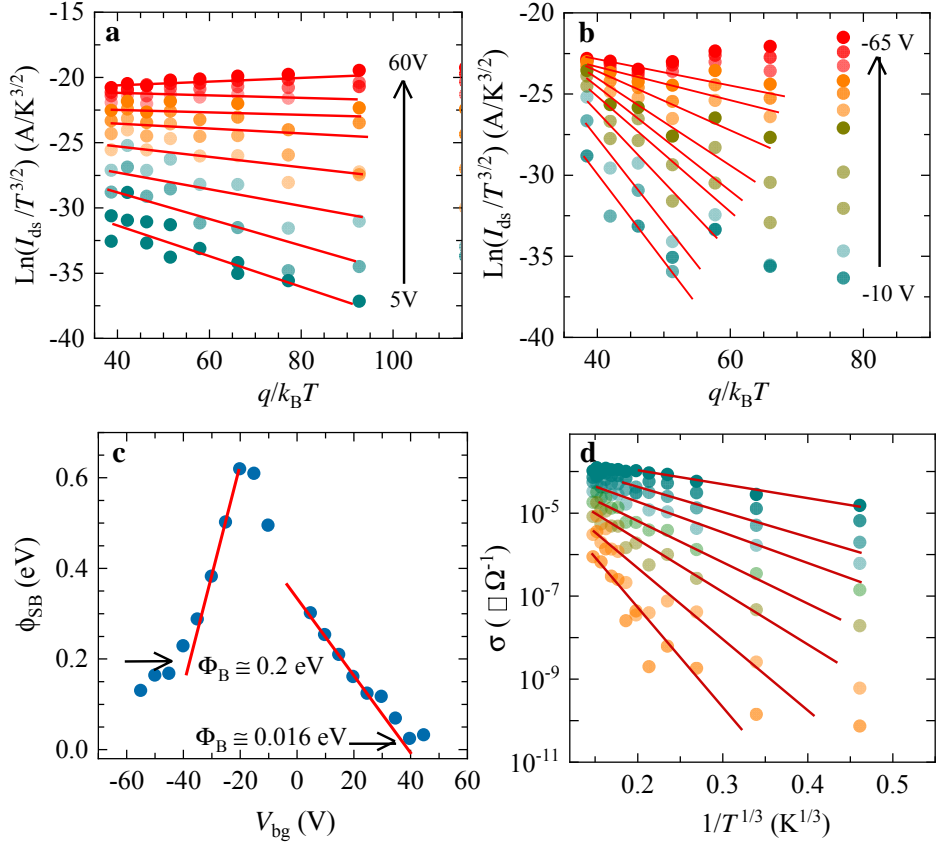

**Figure 4.** (**a**) Drain to source current $I_{ds}$ normalized by a power of the temperature $T$ as a function of charge $q = e$ for the $n = 10$ layers sample and for several positive values of the back gate voltage $V_{bg}$. (**b**) Same as in (**a**) but for negative values of $V_{bg}$. Both data sets in (a) and in (b) were measured under $V_{ds} = 50$ mV. In both panels red lines are linear fits from which we extracted the gate voltage dependence of the Schottky barrier $\phi_{SB}$ ($V_{bg}$) between metallic contacts and the semiconducting channel. (**c**) $\phi_{SB}$ as a function of $V_{bg}$, showing that in the limit of high gate voltages (flat band condition), the extracted Shottky barriers $\Phi_B$ were ~ 200 meV for holes and ~16 meV for electrons respectively. (**d**) Conductivity $\sigma = I_{ds}/V_{ds}\, l/w$, where $l$ and $w$ are length and width of the semiconducting channel respectively, as a function of $1/T^{1/3}$ and for several gate voltages. Gray lines are linear fits indicating that the conductivity as a function of $T$ can be described by the two-dimensional variable range hopping expression.

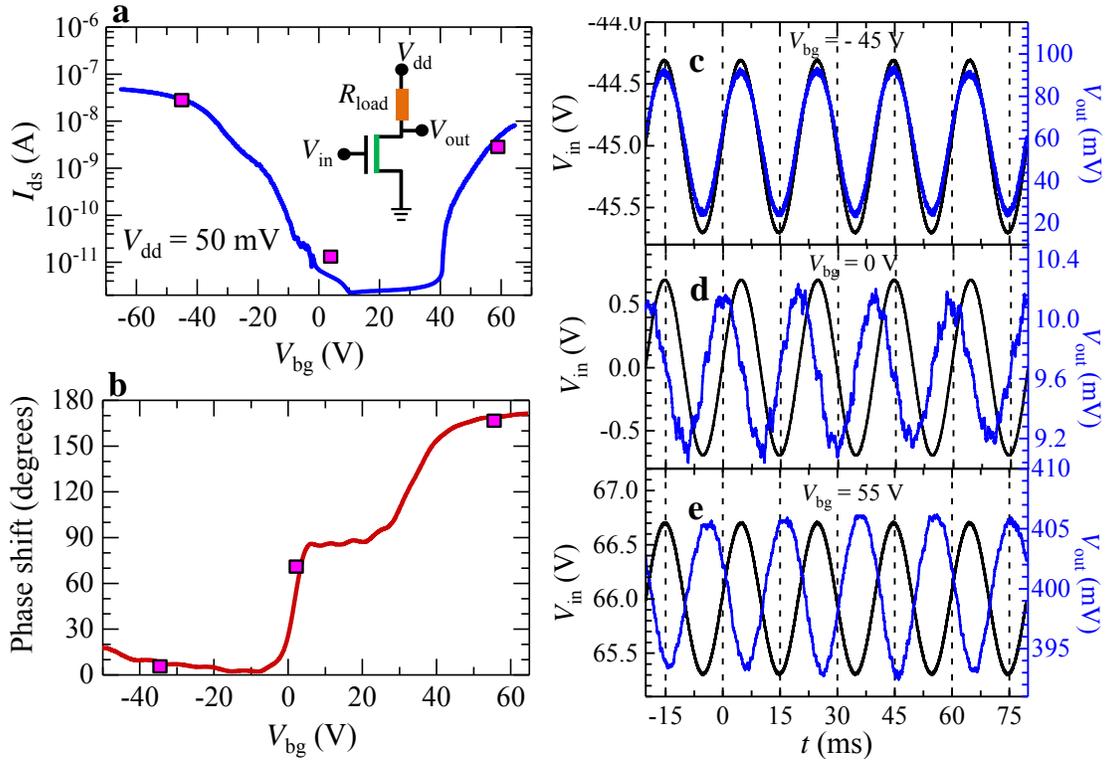

**Figure 5**. Phase-modulation based on a four-layer ambipolar ReSe$_2$ FETs. (**a**) $I_{ds}$ as a function of $V_{bg}$ for a few-layer ReSe$_2$ field effect-transistor at $T = 275$ K. This trace was acquired under a drain supply voltage $V_{dd} = 50$ mV. Inset depicts the scheme of measurements where $R_{load}$ is a load resistor and $V_{dd}$ is the bias voltage. $V_{in-ac}$, which is a superposition of DC and AC (~1.5 V) biases, was applied to the back-gate while $V_{out}$ corresponds to the read-out voltage. Magenta squares indicate the DC back-gate voltages chosen to superimpose an oscillatory AC signal to extract the relative phase-shift between $V_{in}$ and $V_{out}$. (**b**) Relative phase shift as a function of $V_{bg}$. By increasing $V_{bg}$ from negative values we tuned the phase-shift to 90° and then to ~180°. This is clearly illustrated by panels (**c**), (**d**) and (**e**) which display $V_{in}$ (black traces) and $V_{out}$ (blue traces) as a function of time $t$ for various gate voltages.